\documentclass[conference]{IEEEtran}
\IEEEoverridecommandlockouts
\usepackage{cite}
\usepackage{amsmath,amssymb,amsfonts}
\usepackage{textcomp}
\usepackage[ruled, noend]{algorithm2e}
\usepackage{enumitem}
\usepackage{float}
\usepackage{chngpage}
\usepackage{hhline}
\usepackage{booktabs}
\usepackage{multirow}
\usepackage{xcolor,graphicx,float}
\usepackage{hyperref}

\def\BibTeX{{\rm B\kern-.05em{\sc i\kern-.025em b}\kern-.08em
    T\kern-.1667em\lower.7ex\hbox{E}\kern-.125emX}}
\begin{document}

\title{Duet: efficient and scalable hybriD neUral rElation undersTanding*\\
}

\author{
    \IEEEauthorblockN{1\textsuperscript{st} Kaixin Zhang, 2\textsuperscript{nd} Hongzhi Wang, 3\textsuperscript{rd} Yabin Lu, 4\textsuperscript{th} Ziqi Li, 5\textsuperscript{th} Chang Shu, 6\textsuperscript{th} Yu Yan, 7\textsuperscript{th} Donghua Yang}

    \IEEEauthorblockA{\textit{Massive Data Computing Lab, Harbin Institute of Technology}
    \\\ 21B903037@stu.hit.edu.cn, wangzh@hit.edu.cn, {22S003043,7203610517,21S103176}@stu.hit.edu.cn, {yuyan, yang.dh}@hit.edu.cn}
}

\maketitle

\begin{abstract}
Learned cardinality estimation methods have achieved high precision compared to traditional methods. Among learned methods, query-driven approaches have faced the workload drift problem for a long time. Although both data-driven and hybrid methods are proposed to avoid this problem, most of them suffer from high training and estimation costs, limited scalability, instability, and long-tail distribution problems on high-dimensional tables, which seriously affects the practical application of learned cardinality estimators. In this paper, we prove that most of these problems are directly caused by the widely used progressive sampling. We solve this problem by introducing predicate information into the autoregressive model and propose Duet, a stable, efficient, and scalable hybrid method to estimate cardinality directly without sampling or any non-differentiable process, which can not only reduce the inference complexity from $O(n)$ to $O(1)$ compared to Naru and UAE but also achieve higher accuracy on high cardinality and high-dimensional tables. Experimental results show that Duet can achieve all the design goals above and be much more practical. Besides, Duet even has a lower inference cost on CPU than that of most learned methods on GPU.
\end{abstract}


\section{Introduction}
\label{Section: introduction}
Cardinality estimation is a classical problem of the database that aims to estimate how many tuples will be selected by a given query. And it's critical since most RDBMS's query optimizers evaluate the query plan according to the cardinality~\cite{AreWeReady}, so the query optimizer's effectiveness depends on accurate cardinality estimation. Also, during the optimization, the estimator is called for many times to evaluate the candidate query plan, so the estimator must be accurate and efficient. Even though a lot of traditional cardinality estimation methods~\cite{TradCE1, SortingCE, BloomFilterCE, SamplingCE1, SamplingCE2, SamplingCE3, SamplingCE4, SamplingCE5, SamplingCE6, HistogramCE1, HistogramCE2, HistogramCE3, HistogramCE4, HistogramCE5, HistogramCE6, HistogramCE7, HistogramCE8} including bloom-filter-based, histogram-based, and sampling-based methods have been proposed to draw a sketch of data for estimating the queries' selectivity, these methods' error is as high as $2e^5$ when facing the complex query workloads and data distribution~\cite{AreWeReady}. Since cardinality estimation directly affects the effectiveness of query optimization, a more accurate cardinality estimation method is urgently needed.

In recent years, AI-driven cardinality estimation methods~\cite{Naru, NeuroCard} have made great progress in accuracy. These methods can be divided into three classes: \textbf{query-driven}, \textbf{data-driven}, and \textbf{hybrid} methods since the first two methods learn from query and data respectively, and the latter performs hybrid learning from both query and data. However, all these three kinds of methods have their shortcomings, we summarize their most common problems into the following five main problems and discuss them by each kind of method.

\begin{enumerate}[label={(\arabic*)}]
    \item \label{Prob.1}High computation complexity and high memory requirement for complex range queries during inference.
    \item \label{Prob.2}Long-tail distribution problem on high-dimensional tables, i.e., estimator's maximum error is significantly higher than the 99th percentile error.
    \item \label{Prob.3}High computation and memory cost for hybrid training.
    \item \label{Prob.4}Instability that gives uncertain results.
    \item \label{Prob.5}Hard to adapt the workload drift problem.
\end{enumerate}

The query-driven methods are end-to-end models, such as MSCN~\cite{MSCN}, RobustMSCN~\cite{RobustMSCN}, LW-XGB~\cite{LW}, and LW-NN~\cite{LW}, they learn to estimate cardinality from the labeled query directly. Such methods are regression models essentially~\cite{AreWeReady}, and must be re-trained or fine-tuned when the workloads drift to maintain accuracy due to the i.i.d. (independent and identically distributed) assumption of machine learning, which leads to the Problem (5). Although some valuable works~\cite{Warper, RobustMSCN} try to improve such end-to-end methods by designing model-updating-method to solve the workload distribution drift problem, or adding query masking to make the model generalize better on unseen predicates, the workload drift problem is still not completely solved since learning from queries requires the future incoming queries to have independent and identically distribution with the training queries, which cannot be guaranteed. This problem is a fundamental flaw of the query-driven method and can hardly be completely avoided by designing more advanced models. 

The data-driven methods~\cite{Naru, NeuroCard, DeepDB, FLAT, FACE, Pre-train} learn from the table itself, so they are irrelevant to query workloads and can adapt to any query workload. All data-driven methods can be roughly divided into three categories: probability-based~\cite{Naru, NeuroCard, FACE}, SPN-based~\cite{DeepDB, FLAT}, and pre-training-based\cite{Pre-train}. Unfortunately, they each have some shortcomings that limit their applications. For probability-based methods, Naru and Neurocard have Problems (1, 2, 4) since they all introduce progressive sampling into their inference. We will analyze it in \autoref{Subsection: Scalability and Inference Cost Evaluation} and confirm this in experiments of \autoref{fig:scalability}. The efficiency of FACE~\cite{FACE} depends on whether there are similar queries that already have converged buckets. If not, the iteration of sampling would be inefficient (Problem(1)) as the authors stated in the paper~\cite{FACE}. The SPN-based methods perform better in inference speed and scalability but have poorer accuracy than Naru. For example, DeepDB~\cite{DeepDB} suffers from low accuracy (Problem (2)) since it introduces the conditional independence assumption. Although FLAT~\cite{FLAT} has removed this assumption, its accuracy barely reaches a level similar to Naru according to the reported experimental results. The pre-training-based method~\cite{Pre-train} had achieved a very high performance in their experiments, which shows great potential for applying the pre-training model to the CE problem. However, such a method highly relies on the assumption that most real-world datasets have common patterns in the frequency and correlation distributions. It is still unknown to what extent this assumption can be satisfied in the real world. And when meeting a dataset with extremely skewed distribution and complex correlation, the effectiveness of this method still needs to be further studied.

The hybrid methods~\cite{UAE} use both query-driven and data-driven methods to improve the estimation accuracy. Peizhi Wu and Gao Cong proposed a hybrid cardinality estimation method UAE~\cite{UAE} by using the Gumbel-Softmax Trick. They analyzed the gradient flow of Naru in detail and proved that the progressive sample hinders the back-propagation in training. The UAE is the first unified deep autoregressive model to use both data as unsupervised information and query workload as supervised information for learning joint data distribution. By backward propagating gradient on the estimation process, the predicted cardinality can be used as supervisory signals along with the predicted distribution to train the model. However, the Gumbel-Softmax Trick that it uses is still a sampling method, so it still suffers from the Problems (1, 2, 4) of data-driven methods. Besides, the most important problem is the high memory and computation cost of UAE's hybrid training (Problem (3)) caused by using hybrid training and progressive sampling simultaneously. We analyze it in \autoref{Section: Hybrid Training with Query Workloads}.

% In summary, Problems (1-4) are usually caused by the sampling method used in the current data-driven and hybrid methods such as Naru~\cite{Naru}, FACE~\cite{FACE}, and UAE~\cite{UAE}, and Problem (5) is caused by the query-driven methods only learning from specific queries.

In this paper, we propose a novel methodology that introduces the predicate information into the autoregressive model by learning from virtual tables and a hybrid approach (\textbf{Duet}\footnote{Duet's code is available at: \url{https://github.com/GIS-PuppetMaster/Duet}}) for efficient and scalable cardinality estimation. Duet can address the five problems simultaneously by removing the sampling process. Specifically, our approach enables accurate and deterministic cardinality estimation for various range query workloads, high cardinality, and high-dimensional tables, with much lower time and memory costs since no sampling is needed. Furthermore, since Duet's estimation process is differentiable and considering the temporal locality of queries, users can further improve the long-tail distribution problem by applying hybrid training on historical queries.

Our contribution can be summarized as follows.
\begin{itemize}
    \item We proposed a novel methodology for modeling data-driven cardinality estimation which can deal with range queries without sampling and any non-differentiable process during inference. Taking a neural network inference as a basic operation, Duet can achieve O(1) time complexity during inference with only a single network forward pass for each estimation. This feature makes Duet can even be inferred on the CPU with less time cost than that of most learned methods on GPU. Also, by removing the sampling process, Duet achieves higher accuracy on high-dimensional tables significantly.
    
    \item Duet is a differentiable cardinality estimation method and naturally supports hybrid training with higher efficiency. 
    
    \item Benefiting from the sampling-free estimation design, Duet is a stable and practical cardinality estimation method that always gives a determinate result. Compared with other hybrid methods, Duet saves a massive amount of GPU memory and calculation during training.
    
    \item The experimental results demonstrate that Duet can achieve less time cost of inference, lower memory cost, and similar or higher accuracy than data-driven and hybrid cardinality estimation baselines represented by Naru and DeepDB and their derivatives. Compared with all query-driven baselines, Duet also achieves much better accuracy and can adapt to any query workload naturally without fine-tuning.
\end{itemize}

\section{Related Works}
Cardinality estimation is a hot-spot problem that has been studied for decades, the history of this study can be divided into two stages: the traditional method stage and the learned method stage.

\textbf{Traditional Methods.} The traditional methods have been fully researched since the RDBMS is proposed, the common core idea of these methods is to build a sketch for the data. The most representative methods include Sampling~\cite{SamplingCE1, SamplingCE2, SamplingCE3, SamplingCE4, SamplingCE5, SamplingCE6}, Sorting~\cite{SortingCE}, Histograms~\cite{HistogramCE1, HistogramCE2, HistogramCE3, HistogramCE4, HistogramCE5, HistogramCE6, HistogramCE7, HistogramCE8}, and BloomFilter~\cite{BloomFilterCE} ~\cite{ExpCESurvey}. These methods usually have a low accuracy since introducing independent assumptions or losing too much information during sketching. However, a significant advantage of these methods is they do not rely on the query workloads since they learn directly from the data. There are also traditional methods that learn from queries rather than data~\cite{Self-TuningHistograms}, however, such methods are not commonly used in existing DBMS.

\textbf{Query-driven Methods.} As the field of machine learning evolves, the regression methodology of the learned method was used for estimating cardinality~\cite{AreWeReady}. These query-driven methods ~\cite{MSCN, QueryDriven1, QueryDriven2, QueryDriven3, QueryDriven4, QueryDriven5} model cardinality estimation as a regression problem and train machine learning models from certain query workloads. Compared to traditional methods, learning from queries can achieve higher accuracy. However, when the query workloads drift from the training workloads, the model's performance drops severely. Although there are some drift detection and updating algorithms are proposed~\cite{Warper}, this type of method still suffers from workload drift and high training costs. Recently, a novel query-driven method~\cite{RobustMSCN} that uses query masking to improve its robustness was proposed. However, its accuracy improvement relative to MSCN is still smaller than that of other advanced data-driven and hybrid methods relative to MSCN~\cite{RobustMSCN, Pre-train, FLAT, UAE}.

\textbf{Data-driven Methods.} To solve the workload drift problem, a few works use unsupervised machine learning to model the data. Probabilistic graphical models (PGM) ~\cite{BaysianNetwork} use Bayesian networks to model the joint data distribution, however, it also relies on conditional independence assumptions. Other works use kernel density estimation (KDE)~\cite{KDE} to avoid the conditional independence assumptions, but their performance is not satisfying enough. DeepDB~\cite{DeepDB} proposed Relational Sum Product Networks (RSPN) to model the data distribution. Although it still involves the conditional independent assumption, it can achieve better accuracy due to the RSPN design. An SPN-based method FLAT~\cite{FLAT} has removed the conditional independent assumption and achieves relatively higher accuracy. Naru~\cite{Naru} is a pioneering work that has become a cornerstone of many other data-driven and hybrid cardinality estimation methods such as NeuroCard and UAE~\cite{NeuroCard, UAE}. It learns the joint data distribution with the autoregressive model and makes estimations through progressive sampling. But it has limited scalable ability, long-tail distribution on high-dimensional tables, and can't use the information of queries to enhance the training as discussed in \autoref{Section: introduction}. FACE~\cite{FACE} is an autoregressive method based on a flow model. It can remove the progressive sampling and outperforms Naru in scalability, but its efficiency depends on whether there are already converged buckets. If not, it still takes an expensive iteration to sample. The recently proposed pre-training-based method~\cite{Pre-train} shows a very high performance in their experiments, which suggests its great potential. However, such a method highly relies on the assumption that most real-world datasets have common patterns in the frequency and correlation distributions.

\textbf{Hybrid Methods.} In recent years, UAE~\cite{UAE} has proposed modifying the progressive sampling used by Naru into differentiable to achieve hybrid training. By doing so, UAE can outperform Naru in most scenarios. Since UAE still needs to do progressive sampling to deal with range queries, UAE not only suffers from the scalable problem, but it also costs much higher memory and computational costs during training.

\section{Problem Definition}
Consider a relation table $T$ that consists of $N$ columns $\{C_1, C_2, ..., C_N\}$. The number of tuples is $|T|$, the domain region of column $C_i$ is $R_i$, and the number of $C_i$'s distinct values is $d_i$.

\textbf{Query.}
A query is a conjunction of several predicates, and each predicate contains three parts: column $C_i$, predicate operator (a function) $pred_i$, and predicate values $v_i$. Let $pred_i \in \{=, >, <, \geq, \leq\}$, $v_i \in R_i$. 

\textbf{Cardinality.}
Given a query $q$, $Card(q)$ is defined as how many tuples in $T$ satisfy the constraints of the given predicates in $q$. The selectivity of $q$ is defined as $Sel(q) = Card(q)/|T|$.

\textbf{Supported Queries.}
As the definition above shows, Duet can support both equalization predicate and range predicates including but not limited to $>, <, >=, <=$. Since Duet shares the framework of Naru, it also supports joins just like NeuroCard does~\cite{NeuroCard}, which is also based on Naru and learns from the full-out join table to estimate cardinality for join queries. As for disjunction between predicates, its estimation can be performed by converting disjunction into conjunction.

\textbf{Cardinality Estimation.}
Given a query $q$ on table $T$ with columns $\{C_1, C_2, ..., C_N\}$, the selectivity of $q$ can be defined as $$\scriptsize Sel(q)=\sum_{x=(c_1,...,c_N)\in C_1 \times ... \times C_N}{I(x)p(x)}$$, where $p(x)$ is the occurrences frequency of tuple $x$, and $I(x)$ is an indicator function which returns 1 if tuple $x$ is subject to the predicate constraints of query $q$, and returns 0 otherwise. 

\section{Methodology}
The basic idea of our approach is to learn the joint distribution with a virtual table containing predicate conditions rather than the original table, which gives Duet the ability to deal with range queries without sampling. To achieve this goal, we solve the main challenges of the expensive cost of generating or sampling from the virtual table, and the difficulty of dealing with multiple predicates for a single column. In this section, we first overview our approach and then introduce the techniques to address the challenges in the following parts.

\subsection{Overview} 

\begin{figure}[tp]
  \centering
  \includegraphics[width=0.9\linewidth]{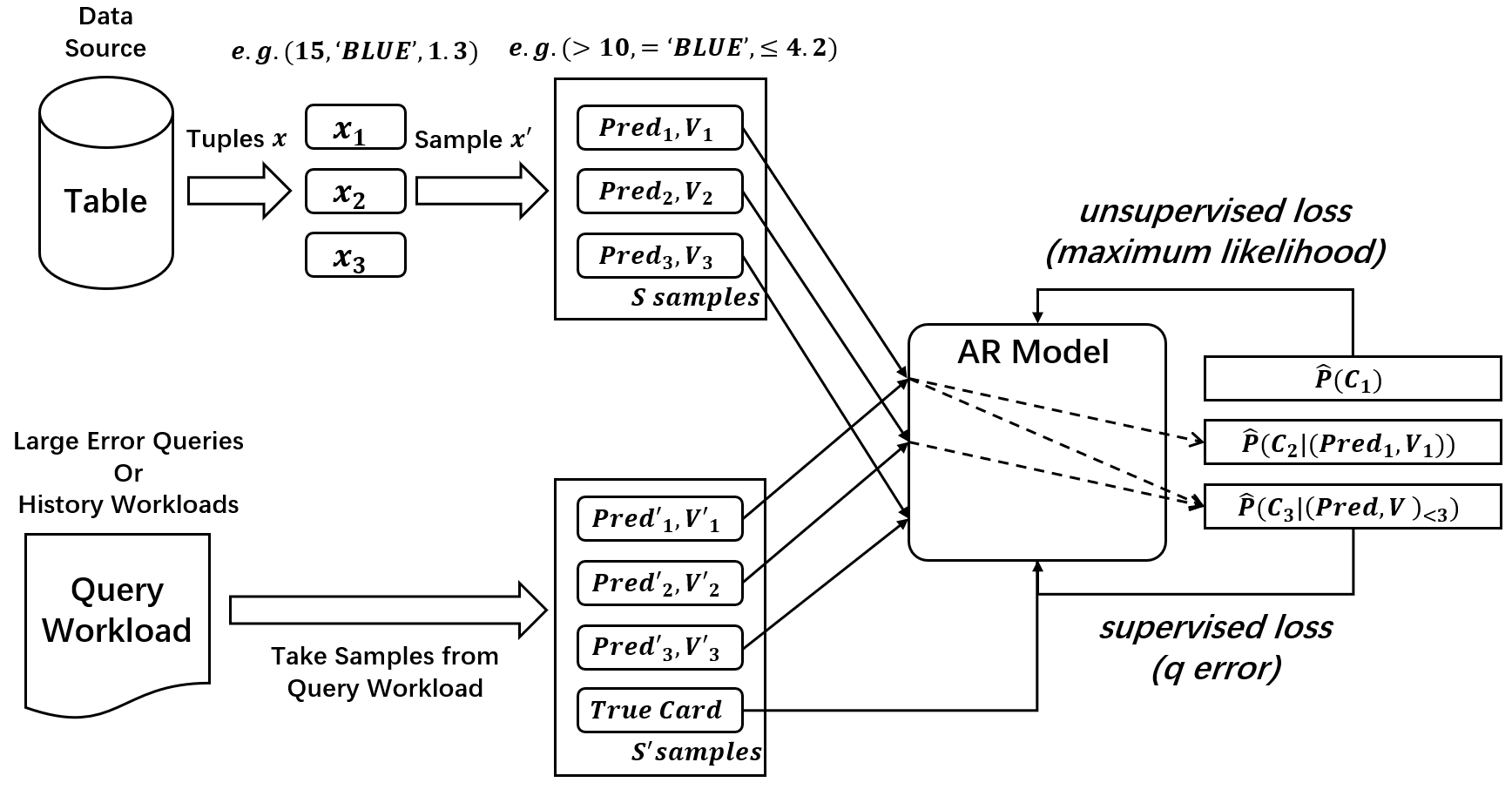}
  \caption{Overview of Duet. Duet learns from both data and queries and makes a differentiable estimation without sampling.}
  % \vspace{-0.2cm}
  \label{fig:Overview}
\end{figure}

\textbf{Motivation \& Challenge.} We aim to address the five main problems introduced in \autoref{Section: introduction}. Firstly, to completely avoid the workload drift problem(Problem(5)), we aim to develop a method that supports both data-driven and hybrid training. Although there is already such a method named UAE~\cite{UAE}, UAE has a high training overhead (Problem(3)) and other problems of Naru since Naru is the basis of UAE. As a method whose cornerstone is to learn the joint distribution of data from tuples with the autoregressive model, which is equivalent to only learning with equivalence predicates, Naru faces three major problems (Problems (1, 2, 4)). The first is \textbf{high computation and memory costs}. Naru uses progressive sampling to deal with range queries, which means the model needs to infer thousands of samples for each step, such computation can only be performed on GPU with limited latency. Also, during one round of the cardinality estimation process, Naru has to infer for $n$ times, where $n$ denotes the number of columns constrained by predicates. So it is still difficult to match the traditional methods' speed in real-world scenarios with hundreds of columns. The second problem is the \textbf{long-tail distribution problem for high-dimensional tables} caused by the error accumulation of progressive sampling\autoref{table:Multiple_Accuracy}. Finally, data-driven methods are \textbf{unstable}, which will give uncertain estimation results for the same queries since the sampling mechanism introduces randomness~\cite{AreWeReady}.

Therefore, the fundamental challenge in solving these issues of existing methods is how to move beyond Naru's autoregressive model of the dataset and develop a novel method to learn the conditional probability distribution that includes predicates to avoid sampling during prediction.

\textbf{Core Idea.} To avoid sampling during inference, we can directly estimate the selectivity of a query by utilizing the predicted conditional probability distribution. As shown in \autoref{fig:Overview}, instead of estimating $p(x)$ as Naru, we directly estimate the selectivity $Sel(q)$ under given predicates. Let $\mathcal{P}_i=(pred_i, v_i)$ denote that column $i$ is subject to the constraint of predicates operator $pred_i$ and the predicates value $v_i$. For example, $(>=, 1.23)$ represents the event that $column_i>=1.23$. With the autoregressive decomposition, we can estimate the selectivity with the following formulation:
%\footnote{We organize the input in $(v_i, pred_i)$ order in our code.}
{\small
\begin{align}
Sel(q) &= p(\mathcal{P}_1, ..., \mathcal{P}_N)\\
&=\prod_{i=1}^{n}{p((pred_{i},v_{i})|(pred,v)_{<i})}\\
&=\prod_{i=1}^{n}{\sum {Pred_i(R_i, v_i)P(C_i|(pred,v)_{<i})}}
\end{align}
}%
, where $Pred_i(\cdot)$ is the predicate operator in the query's predicates, it returns a vector to indicate the distinct values that obey the $i$th predicate's constraint. We would like to emphasize that the learning target of previous works such as Naru and UAE is $P(C_i|v_{<i})$, which does not contain the predicate operator's information like Duet and causes that they can only estimate equivalent predicate through the model and have to use the sampling method to deal with range queries. By introducing predicate information into the autoregressive model, the methodology used by Naru and UAE can be regarded as a special case where the predicate operator in Duet can only take the value '='.

%$E(\cdot)$ is an encoding function that can encode the $pred_i$ and $v_i$ into a vector,
To accomplish such an estimation, Duet learns from a virtual table's tuples that contain predicate information by sampling during training as shown in \autoref{Section: Learn From Virtual Table with Predicates} and \autoref{Section: Training on Virtual Table Efficiently}. As \autoref{fig:Overview} shows, for a given query, Duet only requires a single forward pass of the neural network with a single input, followed by a few simple vector multiplications to obtain the estimated selectivity. This makes Duet faster than existing data-driven methods and gives a deterministic result for a given query. Moreover, since no sampling is involved in this process, gradient information can be backpropagated from the Q-Error. We leverage this property to integrate query-driven methods by simultaneously using unsupervised loss from the data distribution and supervised loss from the query workload during training, as discussed in \autoref{Section: Hybrid Training with Query Workloads}, which enables higher accuracy. Additionally, for queries with large estimation errors during actual use, we can collect them and perform targeted fine-tuning of the model to improve the long-tail distribution problem. Also, we propose MPSN for Duet to support multiple conjunction predicates as introduced in \autoref{Section: Support Multiple Predicates on A Single Column}.

\subsection{Learn From Virtual Table with Predicates}
\label{Section: Learn From Virtual Table with Predicates}
In this subsection, we introduce Duet to directly estimate range queries' selectivity by learning from the virtual table.

We know that for an input tuple $x=(x_1,...,x_N)$ and a model output $\hat{P}(C_1), ...,\hat{P}P(C_N|x_{<N})$, Naru maximizes the likelihood estimation to learn the output distribution. This motivates us to add the range predicate's information into the input sample $x'=(\mathcal{P}_1, ..., \mathcal{P}_N)$, where $x_i$ is subject to the constraints of $\mathcal{P}_i$ and directly estimate the conditional probability distribution $\hat{P}(C_i|(\mathcal{P}_{i-1},..., \mathcal{P}_1)$ through an autoregressive model, denoted as $\hat{P}(C_i|\mathcal{P}_{<i})$. This process is not a just simple change of the model's input but completely changes the problem modeling of the autoregressive learning process. Then we can train Duet with cross-entropy loss and then directly estimate the conditional probability distribution from $\hat{P}(C_i|\mathcal{P}_{<i})$. Such a training process can be understood from two perspectives.

\begin{itemize}
    \item From the meaning of the cross-entropy loss, given a query $x'$, Duet outputs the estimated distribution $\hat{P}(C_i|\mathcal{P}_{<i})$ (denoted by $\hat{p}(x')$), and the ground-truth distribution is $P(C_i|\mathcal{P}_{<i})$ (denoted by $p(x')$). The cross-entropy loss can be written as: $H(\hat{p},p)=-\sum p(x')\log \hat{p}(x')$. Also, since the KL divergence $D_{KL}(p||\hat{p})=\sum_{i=1}^{n}P_i \log \frac{P_i}{\hat{P}_i}=\sum p(x') \log {\frac{p(x')}{\hat{p}(x')}}=H(p,\hat{p})-H(p)$ and $H(p)$ is the data's entropy which is a fixed value, minimizing the cross-entropy equals to minimize the KL divergence between $p(x')$ and $\hat{p}(x')$. Minimizing KL divergence will shrink the distance between the two distributions.

    \item From the meaning of maximum likelihood estimation, the maximum likelihood estimation of Duet aims to find a $\hat{p(x')}$ that fits the ground-truth distribution $p(x')$ best. Therefore, Duet performs maximum likelihood estimation whose target is maximizing $loss(x')=\sum{\log{\hat{p}(x')}}$. Compared with the formula of KL divergence, since the $p(x')$ is a determined value for a given dataset and predicates, the maximum likelihood estimation equals minimizing the cross-entropy and the KL divergence.
\end{itemize}

\begin{figure}[tp]
  \centering
  \includegraphics[width=\linewidth]{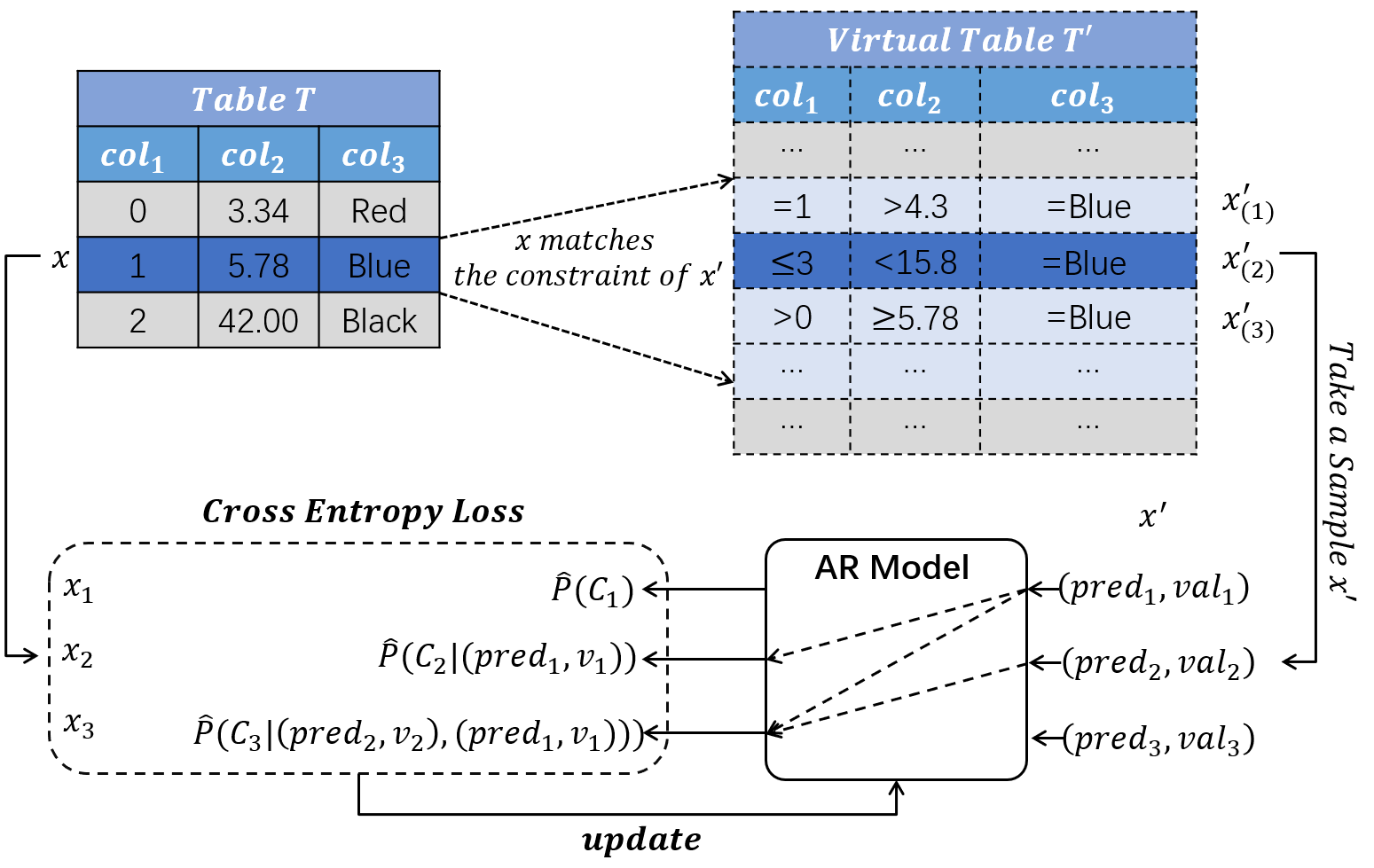}
  \vspace{-0.8cm}
  \caption{The unsupervised training part of Duet. We input a sample $x'$ from the virtual table and use the corresponding $x$ as the label. By updating with cross-entropy, we minimize the KL divergence and make Duet learn the actual data distribution under the constraint of the input virtual tuple $x'$.}
  % \vspace{-0.5cm}
  \label{fig:VirtualTable}
\end{figure}

In summary, as shown in \autoref{fig:VirtualTable}, we give the formal definition of the virtual table. Let $x=(x_1,...,x_N)$ be a tuple of table $T$ and $x' =(\mathcal{P}_1, ..., \mathcal{P}_N)$ be a series of predicates, each contains a predicate operator and a predicate value. If $x_i$ satisfies the constraint of $\mathcal{P}_i$, we denote it as $\mathcal{S}(x_i,\mathcal{P}_i)=1$, otherwise $\mathcal{S}(x_i,\mathcal{P}_i)=0$. Then We define a function $I(\cdot)$ as follow:
\begin{align*}
\begin{split}
I(x, x')= \left \{
\begin{array}{ll}
    1,     & \forall i\in [1,n], x_i \in x, \mathcal{P}_i \in x', \mathcal{S}(x_i,\mathcal{P}_i)=1\\
    0,     & otherwise
\end{array}
\right.
\end{split}
\end{align*}
, and the virtual table $T'$ is defined as:
$$T'=\{x'| \forall x \in T, I(x, x')=1\}$$

% for a tuple $x=(x_1,...,x_N)$ of table $T$, a set of virtual tuples $x'=(x'_1,...,x'_N)$ contains range predicates that make tuple $x$ satisfy their predicate constraints. These virtual tuples make up a virtual table $T'$. 

Duet draws a sample $x'$ from $T'$ as input and outputs a probability distribution of each column's distinct values in Table $T$. Therefore, the learning target of Duet is to give a virtual tuple $x'$, and the model learns a conditional probability distribution matching the actual data distribution under the constraint of predicates in the input virtual tuple $x'$. This is achieved by the KL divergence loss function that can be simplified to the cross-entropy loss.

% Thirdly, since the sampling processing is removed, the estimation now is differentiable so Duet can naturally achieve hybrid training at a low cost.

%We can also pre-generate some query workloads to make a more efficient convergence during the first time of hybrid training.

% }

\subsection{Training on Virtual Table Efficiently}
\label{Section: Training on Virtual Table Efficiently}
In this subsection, we introduce how to train Duet efficiently by dynamically generating virtual tuples during training.

As \autoref{Section: Learn From Virtual Table with Predicates} discussed, the virtual table $T'$ is generated from the original table $T$ with the following rule satisfied. For a tuple $x=(x_1,...,x_N)$ of table $T$, all virtual tuples $Set(x')$ that make tuple $x$ satisfy their predicate constraints can be generated. However, if the column $col_i$ is in categorical type with a massive amount of distinct values, the cost of generating such a virtual table is unacceptable since the cardinality of the virtual table can reach up to $\prod_N{k*d_i}$, where $k$ is the number of predicate types and $d_i$ is the number of distinct values of column $i$. Also, since we require the label $x$ to satisfy the predicates of the model's input virtual tuple $x'$, even if we can pre-generate the virtual table, finding the corresponding $x$ of a given $x'$ would also cost much computation resource and time. Therefore, instead of generating the virtual table and searching for the corresponding $x$, we use a uniform sampling method to dynamically generate the virtual tuple $x'$ with the randomly selected tuple $x$ during stochastic gradient descent.

Intuitively, since $x$ is sampled from table $T$, we only need to generate a predicate $(pred_i, v_i)$ for each column of $x$ that makes $x_i$ satisfies it, then the $x'$ that consists of these predicates would be satisfied by at least one tuple in the original table $T$. For example, as shown in \autoref{fig:Overview}, for the target value $col_1=1$, we can draw a sample $(pred_i='\leq', v_i=1.3)$, $v_i$ can take any values larger or equal to 1.3 within $R_i$. For the component $x_i$ of each column $i$ in $x$, we randomly select a predicate operator for this column. Assuming that $r_i$ is the range that makes $x_i$ satisfy the predicate operator, we use uniform sampling to select a $x'_i$ from $r_i$ as the predicate value.

We observe that such tuple-by-tuple sampling is too expensive. To reduce the sampling process overhead furthermore, we implement a vectorized GPU version based on the \textit{collect\_fn} of Torch's Dataloader, it can process a batch of $x_i$ at one time. Not like the sampling algorithm used by Naru~\cite{Naru} and UAE~\cite{UAE} during inference, our sampling algorithm draws tuples uniformly and ignores data skewness in the data. This is because UAE has to estimate the probability of the range query predicates by sampling the probability of the equivalent predicates, while the distribution of the target (distinct value of each column) is determined and known. In this case, there is a data skew problem and their work's solution is to use progressive sampling instead of uniform sampling. In Duet, the objects we sample during the training process are predicates in the virtual table rather than tuples in the data, and the distribution of predicates in future queries is usually uncertain and unknowable. In order to consider the robustness of the method, this paper designs from the worst possible situation (i.e., the future query is completely random, and the distribution is unknown). In this case, using uniform sampling will be the most reasonable method, because it ensures that each predicate is trained evenly. However, in real-world scenarios with strong query time locality, it's possible to use the historical queries' distributions to guide the sampling by replacing the uniform sampling with importance sampling.

% $(upper-lower)*r+lower$ in
Our sampling algorithm is shown in \autoref{Algorithm: Virtual Tuple Sampling Method}. Specifically, we take all column values $tuples\_i$ within a batch as a vector and assign a predicate operator for each value as shown in $Line\ 22$. Since the advanced indexing of PyTorch makes a deep copy of data and leads to high costs compared to the slice operator which does no copy, we shuffle the training data at each epoch, divide the data batch into slices, and assign different predicate operators for each slice in $Line\ 7$ instead of randomly assigning predicate operators. Then, for each slice, we calculate the lower and upper sampling boundary of the index of this column's $dvs$ in $Line\ 12-14$. Finally, we generate a batch of float numbers $r$ belonging to $[0,1)$ and calculate each sample value with the formula in $Line\ 17$. Since each column's sampling process is independent for each column, we can sample in parallel with a high speedup ratio. We implement this algorithm with Cpp-Extension and use multi-threading to parallelize it to avoid the Python GIL limitation. During the implementation, we observed that the $Tensor.put\_index\_$ function of $LibTorch\ 2.0.1$ on GPU is $1000\times$ slower than its CPU version, so the whole sampling algorithm runs on CPU with calling of $Tensor.put\_index\_$ as less as possible since this function is the performance bottle-neck of this algorithm.

To accelerate the convergence of training, we replicate the data batch of $tuples\_i$ for $\mu$ times (usually takes the value of 4) and sample the corresponding virtual tuples independently as shown in $Line\ 21$. Thus, each tuple can be trained for $\mu$ times with different predicates within a gradient descent step, and the model's performance won't suffer from a large batch size.

\begin{algorithm}[htbp]
    \caption{Parallel Vectorized Sampling}%算法名字
    \label{Algorithm: Virtual Tuple Sampling Method}
    \LinesNumbered %要求显示行号
    \scriptsize
    \KwIn{data batch $tuples$, batch size $bs$, column number $N$, number of involved predicates' kinds $k$, expand coefficient $\mu$}%输入参数
    \KwOut{$new\_tuples, new\_preds$}%输出
    \SetKwFunction{FMain}{SamplingPerColumn}
    \SetKwProg{Fn}{Function}{:}{}
    \Fn{\FMain{tuples\_i}}{
        slices $\longleftarrow$ DivideDataBatch(tuples, k);\\
        new\_tuples\_i $\longleftarrow$ Initialize with -1;\\
        masks, lower, upper, new\_pred\_i $\longleftarrow$ Initialize with 0;\\
        \ForEach{$s, e$ in $slices$}{
            // $=$, $>$, $<$, $>=$, $<=$ are numbered as 0-5\\
            pred $\longleftarrow$ Randomly assign predicates for slices without repetition;\\
            \uIf{pred==0}{
                new\_tuple\_i[s:e] $\longleftarrow$ tuples\_i[s:e];\\
                new\_pred\_i[s:e, pred] $\longleftarrow$ 1;\\
            }
            \uElse{
                Calculate $lower\_bound$, $upper\_bound$;\\
                masks[s:e] $\longleftarrow$ lower\_bound $<$ upper\_bound;\\
                lower[s:e], upper[s:e] $\longleftarrow$ lower\_bound, upper\_bound;\\
                new\_pred\_i[s:e, pred][masks[s:e]] $\longleftarrow$ 1;\\
            }
        }
        samples $\longleftarrow$ (upper - lower) * \textit{rand}(bs*f) + lower;\\
        new\_tuple\_i[s:e][masks] $\longleftarrow$ samples[masks];\\
        res $\longleftarrow$ concatenate([new\_tuple\_i, new\_pred\_i], dim=-1);\\
        \Return res;\\
    }
    Replicate $tuples$ for $\mu$ times, so we sample $\mu$ predicates from the virtual table within one epoch for each tuple;\\
    \For{$i\in [0, N)$}{
        Launch threads to run $SamplingPerColumn(tuples[...,i]);$\\
    }
    new\_tuples, new\_preds $\longleftarrow$ Collect results from threads;\\
    \Return concatenate(new\_tuples,dim=-1), concatenate(new\_preds,dim=-1);\\    
\end{algorithm}

\textbf{Encoding.} We use binary encoding for the predicate value $v_i$ and one-hot encoding for the predicate operator $pred_i$. For each column's predicate, we concatenate $(v_i, pred_i)$ as its encoding. For the column without predicates, we use the wildcard-skipping techniques proposed by Naru~\cite{Naru} to fill the $v_i$ and set the $pred_i$ vector as a zero vector. For large distinct value columns, we also use embedding to encode $v_i$. These techniques are consistent during training and inference.

\subsection{Hybrid Training with Query Workloads}
\label{Section: Hybrid Training with Query Workloads}

In this subsection, we introduce our solution to UAE's high cost of hybrid training.

Let $s$ be the number of samples and $bs$ be the batch size, since UAE still needs to sample for range predicates' cardinality estimation, and all the gradients of these samples have to be tracked, the actual batch size that UAE uses for query-driven training is $bs\times\ s$. During our evaluation, for training with $bs=2048$ and $s=2000$ on the DMV dataset, which is the parameter used by Naru, it cannot be trained on a 48 GB GPU memory environment due to the out-of-memory exception. This method used a much smaller $s$ and $bs$. However, a small $bs$ will significantly slow the training process, and a small $s$ will lower the precision of cardinality estimation for range queries and further affect the model accuracy via $\mathcal{L}_{query}$.

% Since methods such as Naru~\cite{Naru} and UAE~\cite{UAE} suffer from the long-tail distribution problem caused by the sampling error, especially for high-dimensional tables, we aim to propose a method that combines query-driven and data-driven approaches, so we can use the queries with large estimation errors to fine-tune the model and reduce the max Q-Error. The existing techniques such as UAE~\cite{UAE} use Gumbel-Softmax Trick to improve the sampling process and make the gradient can be backward passed through the estimation process.

Benefiting from the direct modeling and estimation, Duet does not involve any non-differentiable sampling and can perform hybrid training naturally. This allows us to use the supervised signal of queries to improve the model's accuracy. Also, such a mechanism allows users to fine-tune the model based on history query workloads after it is deployed.

Since hybrid training can utilize both types of information to learn the joint data distribution in a single model~\cite{UAE}, we pre-generate a few training workloads and use these workloads for hybrid training to improve the model's accuracy. Note that these training workloads can be real-world history query workloads in real work scenarios. Our loss function consists of two parts: the unsupervised loss $\mathcal{L}_{data}$ and the supervised loss $\mathcal{L}_{query}$. We use Cross-Entropy for $\mathcal{L}_{data}$ and a mapped Q-Error (the definition of Q-Error is the same with UAE~\cite{UAE}) for $\mathcal{L}_{query}$.

How to combine these two parts of loss is critical to the model's accuracy. On the one hand, as we learned from the previous works~\cite{Naru, NeuroCard}, the $\mathcal{L}_{data}$ is workload-independent so minimizing it with higher weights can improve the model in generalization ability. On the other hand, $\mathcal{L}_{query}$ is an effective supplement to help the model achieve higher accuracy. However, it is workload-related, so its application to high weight on $\mathcal{L}_{query}$ will cause the model to degenerate into a query-driven approach and damage its generalization ability. Therefore, we attempt to design a hybrid loss function to make the hybrid training both effective and generalizable.

\begin{figure}[htp]
  \centering
  \vspace{-0.3cm}
  \includegraphics[width=0.9\linewidth]{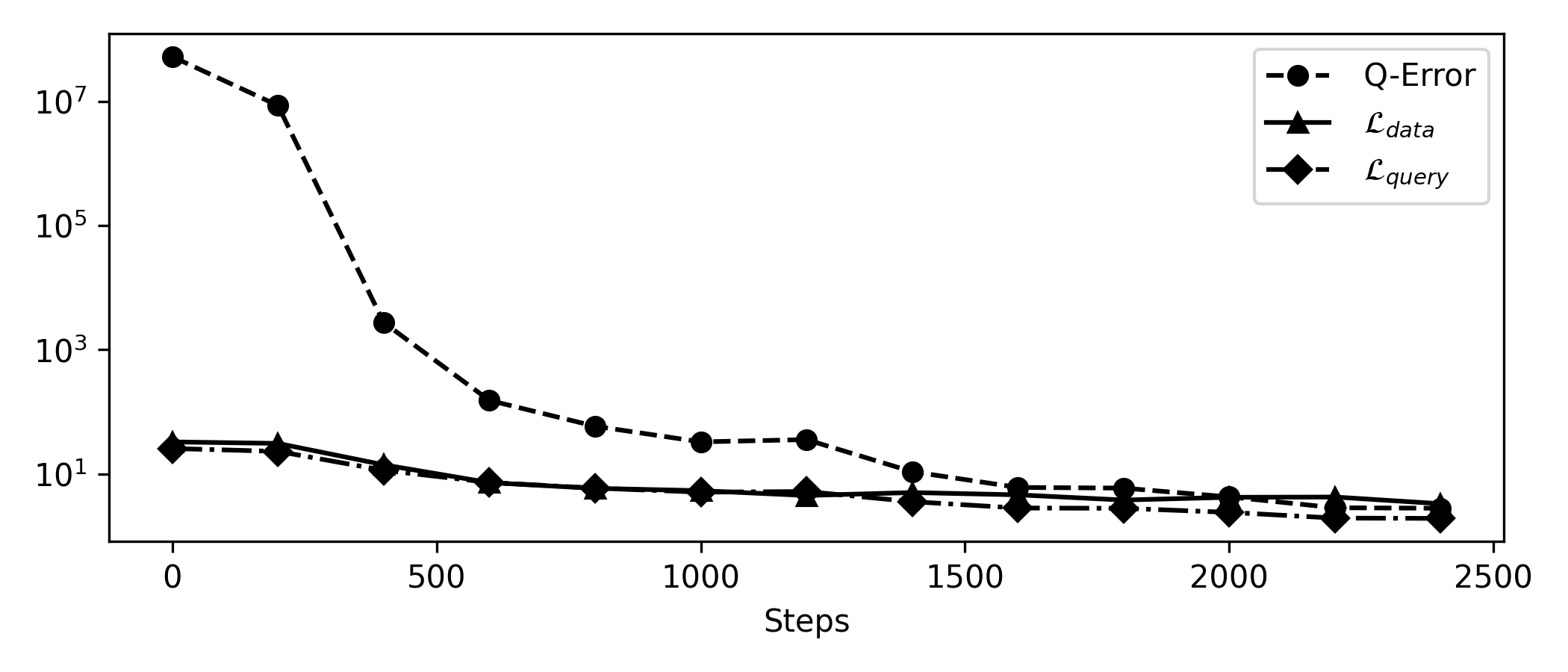}
  \vspace{-0.6cm}
  \caption{The convergence of the Q-Error, evaluated on DMV dataset.}
  \vspace{-0.3cm}
  \label{fig:loss_convergence}
\end{figure}

We observe that $\mathcal{L}_{query}$ takes a huge maximum value and decreases significantly faster than $\mathcal{L}_{data}$ as shown in \autoref{fig:loss_convergence}. Therefore, we use a $log_{2}$ function to map the Q-Error into a much smaller and smooth space to prevent the gradient explosion and unstable training. Since the $log_{2}$ function has a steep gradient in range $[0,1)$, we plus one with the Q-Error to help the training converge stably. As shown in \autoref{fig:loss_convergence}, the improved $\mathcal{L}_{query}$ has similar order and convergence rate with $\mathcal{L}_{data}$. Then, we multiply it by a trade-off coefficient $\lambda$ so that its value is scaled to an appropriate scale, which can improve the model accuracy without affecting the model generalization performance. According to the hyper-parameter study in \autoref{Subsection: Hyper-Parameter Studies}, the $\lambda$ can take a value of $0.1$ in most cases, and it can be adjusted flexibly if this method is used in scenarios where the query workload is relatively stable or extremely unstable to further improve the accuracy. The hybrid loss proposed is shown as follows: $$\mathcal{L}=\mathcal{L}_{data}+ \lambda  \mathcal{L}_{query}=\mathcal{L}_{data}+ \lambda  log_{2}(\text{Q-Error}+1)$$

The detailed hybrid training algorithm is given in \autoref{Algorithm: Hybrid Training Process}. During the training iteration of the mini-batch SGD, we first use the \autoref{Algorithm: Virtual Tuple Sampling Method} to sample a batch of virtual tuples in $Line\ 3$. Then we collect $bs$ queries from training workload $\mathcal{Q}$ in $Line\ 4$. After calculating the loss of both data and queries in $Line\ 5-7$, we perform SGD on $\mathcal{L}$ in $Line\ 8$.

\begin{algorithm}[htbp]
    \caption{Hybrid Training Process}%算法名字
    \label{Algorithm: Hybrid Training Process}
    \LinesNumbered %要求显示行号
    \scriptsize
    \KwIn{dataset $\mathcal{D}$, training workload $\mathcal{Q}$, batch size $bs$, expand coefficient $\mu$, trade-off coefficient $\lambda$}%输入参数
    \KwOut{Trained Model parameters $\theta$}%输出
    Randomly initialize model weights $\theta$;\\
    \ForEach{$batch$ in $\mathcal{D}$}{
        $tuples \longleftarrow $ Sampling through \autoref{Algorithm: Virtual Tuple Sampling Method} with expand coefficient $\mu$;\\ 
        $workloads \longleftarrow$ Collecting $bs$ queries from $\mathcal{Q}$;\\
        $\mathcal{L}_{data} \longleftarrow$ Run unsupervised training on $tuples$;\\
        \textit{Q-Error} $\longleftarrow$ Run estimation directly on $workloads$;\\
        $\mathcal{L}_{query} \longleftarrow log_{2}$ \textit{(Q-Error+1)};\\
        Perform SGD with $\mathcal{L}=\mathcal{L}_{data} + \lambda  \mathcal{L}_{query}$ to update $\theta$;\\
    }
    \Return $\theta$;\\
\end{algorithm}

\subsection{Efficient And Scalable Estimation without Sampling}
\label{Efficient And Scalable Estimation without Sampling}
In this section, we introduce how Duet estimates cardinality for queries (especially for range queries) and discuss the complexity of Duet.

A major part of the existing data-driven and hybrid methods such as Naru, and UAE learn from the original table which is equivalent to only learning with equivalence predicates. To deal with range queries, they all use progressive sampling to approximate the result. Such a solution leads to four major problems (Problems (1-4)) as mentioned in \autoref{Section: introduction}.

% 11/6日：这里新删除了一大段，对Naru分析过于细致占用太多篇幅
% We denote the number of columns that involve at least one predicate to be $n$.
% To alleviate this problem, Naru uses a technique called wildcard-skipping. By learning an embedding vector for each column that represents there are no predicates on the corresponding column during training, all the unnecessary inferences are skipped. However, Some complex queries have even hundreds of columns that involve at least one predicate, , so the problem is still unsolved completely.
First of all, real-world datasets and workloads usually involve multiple columns and predicates. The progressive sampling process will cause $n$ times non-parallelizable model inference (forward pass). This makes those works unavailable in most OLTP scenarios~\cite{AreWeReady, CEBenchMark}. For instance, survey~\cite{AreWeReady} evaluates existing learned cardinality estimation methods using queries involving up to 100 columns. Also, progressive sampling requires inferences with more than 2k samples for one round of estimation~\cite{Naru}, such a calculation has to be accomplished by GPU within a limited time. Clearly, the cost of building a GPU database server cluster is much higher than the ordinary cluster.

Secondly, range queries' cardinality estimation by sampling is a lossy one. Although the authors of Naru prove that progressive sampling is an unbiased sampling and has higher efficiency than uniform sampling, it is still not as good as direct estimation by the learned model for high-dimensional tables, which is demonstrated in the \autoref{Subsection: Accuracy Comparison}. This is because progressive sampling causes an error accumulation during estimating queries on high-dimensional tables since the limited samples can only cover a small area of the high-dimensional hyper-rectangle of range predicates.

Thirdly, as discussed in \autoref{Section: introduction}, each sample of the hybrid training batch needs to be sampled over 2k times, which leads to a high GPU cost for hybrid methods such as UAE~\cite{UAE}.

Finally, sampling during inference introduces randomness into estimation, which leads to indeterministic results for the same queries. This will make it harder for the DBA to locate the potential problems of optimization.

Duet solves these problems by learning from the virtual table with predicate information. Firstly, compared to Naru, taking a neural network inference as a basic operation, Duet reduces the time complexity of range queries from $O(n)$ to $O(1)$ and reduces $s$ times memory, since no sampling process is needed for range queries. As \autoref{Algorithm: Inference for Estimation} shows, Duet estimates the range queries' cardinality with only one time of inference. We first encode the predicates in $Line\ 1$. Then we obtain the conditional probability distributions through a single forward pass of the model in $Line\ 2$. After zero-out each probability with the $Pred_i(R_i, v_i)$ masks in $Line\ 3$ (where $R_i$ is the domain range of column $i$), we finally perform a multiplication of the summary of these probability distributions in $Line\ 4$. Let one forward pass be a basic computation of estimation, Duet's computation complexity is $O(1)$ rather than previous works' $O(n)$. In terms of computation cost, each inference of Duet only requires one single input. Thus, the inference stage can be applied to the CPU.

For the complexity of taking multiplication as the basic computation, without loss of generality, we consider a MADE network with $m$ layers and $h$ hidden units for each layer. The complexity of computing all hidden units is $mh^3$ (the mask can be pre-merged with the weight matrix). Let the encoding length of each column's predicate value be $e$ and the number of columns is $N$, the total input length of Duet is $(e+5)N$, and the complexity of the input layer and output layer is $\mathcal{C}=2((e+5)N)^2h$. So in general, the total multiplication times are $2((e+5)N)^2h+mh^3$, which belongs to $O(N^2)$. \textbf{Note that the $N^2$ part comes from the matrix multiplication and can be highly paralleled on the CPU with AVX instruction set or GPU.} As a comparison, let the number of progressive samples be $s$, and the total multiplication times of Naru and UAE are $s(n-1)\mathcal{C}$, which costs $s(n-1)$ times of multiplication than Duet. 

Secondly, by removing the sampling processing, the sampling error accumulation is removed accordingly. Thus, the long-tail distribution problem caused by the high-dimensional table is alleviated. 

Thirdly, since range queries' cardinality can be predicted without any non-differentiable process, we have a chance to introduce the Q-Error as a feedback signal to the loss function to achieve hybrid training with minimum cost. Considering the temporal locality of real-world workloads, users can use historical queries for hybrid training to improve model accuracy.

Fourthly, as for stability, Duet is stable since no randomness exists during estimation. For a given query, Duet will provide a deterministic result that could help the query optimizer work better and help the DBA to find problems easier.

% Lastly, since Duet directly estimates each column's probability distribution under the given range predicates' condition, it achieves higher accuracy than previous work.

% Here $n$ and $s$ denote the number of columns constrained by the predicate and the progressive sampling amount in Naru, respectively.

% Such a methodology solves a few problems mentioned in \autoref{Section: introduction}.  }

\begin{algorithm}[htbp]
    \caption{Inference for Estimation}%算法名字
    \label{Algorithm: Inference for Estimation}
    \LinesNumbered %要求显示行号
    \scriptsize
    \KwIn{Query predicates $preds$, Zero-out mask $Pred_i(R_i, v_i)$, model $\mathcal{M}(\theta)$ with parameter $\theta$ }%输入参数
    \KwOut{Selectivity $\hat{p}$}%输出
    $inp \longleftarrow$ Encoding predicates $preds$;\\
    $\hat{P}(C_0), ...,\hat{P}(C_N|(pred, v)_{<N})) \longleftarrow$ Forward pass through $\mathcal{M}(\theta)$;\\
    Zero-out each probability in $\hat{P}$ which doesn't satisfy corresponding predicate;\\
    $\hat{p} = \prod{\sum{\hat{P}(C_i|(pred, v)_{<i})}}$;\\
    \Return $\hat{p}$;\\
\end{algorithm}
\vspace{-0.7cm}

\subsection{Support Multiple Predicates on A Single Column}
\label{Section: Support Multiple Predicates on A Single Column}
We focus on the situation that each column has one single predicate at most in previous subsections. In this subsection, we generalize Duet to support multiple predicates on a single column by introducing Multiple Predicates Supporting Networks (\textbf{MPSN}) into Duet to embed multiple predicates.

Benefit from the flexible methodology of Duet, we achieve this goal by expanding the input of Duet from a series of $(pred_i,v_i)$ to $((pred^{1}_i,v^{1}_i),...,(pred^{j}_i,v^{j}_i))$. The conditional part of the conditional probability distribution output by the model has also changed accordingly. During sampling $x'$, we randomly choose the number of the predicates for column $i$.

The key problem is how to combine the variable-length predicates' information to fit the input size of the autoregressive network, which is a problem that Naru does not need to deal with since it uses progressive sampling and deals with multiple predicates' information at the batch dimension. Due to the limited number of predicates, we encode the predicate operator with the one-hot encoder and provide binary, one-hot, and embedding strategies for the predicate values just like Naru. To integrate such variable-length information before inputting them into the autoregressive model, we propose three MPSN candidates to embed the predicates into a fixed-length dummy vector space and evaluate their performance with experiments, as are introduced as follows. Note that each column corresponds to an independent MPSN. 

\textbf{MLP \& Vector Sum.} Since the number of predicates on a single column may not be too large in most queries, we use a simple MLP network to embed each pair of $(pred^{j}_i,v^{j}_i)$ into vectors in parallel, then we sum these vectors up and obtain the final embedding result with a fixed length.

\textbf{Recurrent Network.} The RNN network is a common method for processing variable-length information. We use a simple and shallow LSTM network to embed the predicates. The output of the LSTM is processed by a fully connected layer, and the outputs of different predicates are summed together as the final result.

\textbf{Recursive Network.} We also evaluate the recursive network's performance for embedding. We build a recursive network whose forward pass formulae is $out = MLP(E(pred^{j}_i,v^{j}_i) || out)$, where $E$ is the encoding function, and the default value of $out$ takes zero for the single predicate situation. Compared to the RNN network, the recursive network has no long-distance information forgetting problem. However, with the increasing number of predicates, the calculation costs increase fast. 

Note that only MLP MPSN is order-irrelevant thus we believe that it is the most hopeful structure of MPSN. Thus, we develop a strategy to accelerate the multiple MLP MPSNs without accuracy loss.

\textbf{Parallel Acceleration for MLP MPSN.} Although Duet is not required to predict column-by-column like Naru and UAE, it needs to embed the predicates and process them with MPSN column-by-column. One straightforward solution is to infer in parallel with multi-threading. Since the encoding and MPSNs are independent of each other, the parallelism and speedup would be high. Furthermore, considering the thread-creating cost, we develop a more efficient method to accelerate the MLP MPSN inferring process by merging all MPSNs of all columns into one MLP model with masked weight. Specifically, we set all MLP MPSNs with the same number of layers and the same activation functions, then we merge each fully connected layer by mapping their weight matrix on the main diagonal of a bigger zero matrix and merging their bias vector together as a new ensemble layer. We have tested this method on tables with 100 columns, and it can finish the computation of MPSN much faster. For some extremely high-dimensional tables, this method can reduce the number of inferences of MPSNs by $100\times$ at least.

\section{Experiment Results}

\subsection{Expeimental Settings}

\subsubsection{\textbf{Datasets}} Followed by Naru~\cite{Naru} and UAE~\cite{UAE}, we take the DMV, Kddcup98, and Census to evaluate Duet's performance.
%on high cardinality tables, high-dimensional tables, and high NDV (number of distinct values) tables separately.
\begin{itemize}
    \item \textbf{DMV.}~\cite{DMV} This dataset is real-world vehicle registration information in New York. We follow the preprocessing of Naru~\cite{Naru} and obtain a table with 12,370,355 tuples and 11 columns. These columns include the categorical type and date type. The NDV's (Number of Distinct Values) range is from 2 to 2,774. We use it to evaluate Duet's performance on high cardinality and large NDV tables.
    
    \item \textbf{Kddcup98.}~\cite{Cpu98AndCensus} This dataset is from the KDD Cup 98 Challenge and contains 100 columns and 95,412 tuples with NDV ranging from 2 to 57. This dataset is obtained from UAE~\cite{UAE} and is mainly used to evaluate the scalability of Duet in large column number scenarios. This dataset is used to evaluate Duet's performance on high-dimensional tables.
    \item \textbf{Census.}~\cite{Cpu98AndCensus} Same as UAE~\cite{UAE}, this dataset contains 48,842 tuples and 14 columns with NDV ranging from 2 to 123. This dataset is used to evaluate Duet's performance on small tables.
\end{itemize}

\subsubsection{\textbf{Query Workloads}}
\label{Subsubsection: Query Workloads}
We followed previous works~\cite{Naru, AreWeReady} to generate workloads. Specifically, we sample a tuple from the table and then generate predicates based on them. This method can generate a uniform distribution of query workload~\cite{AreWeReady} that includes a wide range of selectivity.

\textbf{Training Workloads.} We use a random seed 42 to generate a workload that contains $1e^5$ queries as the training workload. To simulate the real-world workload distribution, we take a similar approach to the UAE. We randomly choose a large enough column and sample 1\% of its distinct values as a bounded column, so the model will only be trained on limited predicates. Also, the number of predicates is selected randomly according to gamma distribution to simulate the skew of real-world queries. For the Duet model that is trained without workloads, we denote it by \textbf{Duet$_{\mathcal{D}}$}.

\textbf{Testing Workloads.} We generate two testing workloads and each contains 2k queries for each dataset with the random seeds 1,234 and 42 (the same as training workloads), denoted by \textbf{$Rand-Q$} (Random Queries) and \textbf{$In-Q$} (In-workload Queries), respectively. The former is generated without a bounded column and the number of predicates is selected uniformly. It is used to assess the performance of random queries which reflects the worst-case model performance that the incoming queries are irrelevant to training workloads. The latter is generated in the same way as training workloads and used to evaluate in-workload performance which reflects model performance under realistic queries with temporal locality. With these two different workloads, we can evaluate the influence of whether the training workload overlaps with the testing workload.

\begin{figure}[htp]
  \centering
  \includegraphics[width=\linewidth]{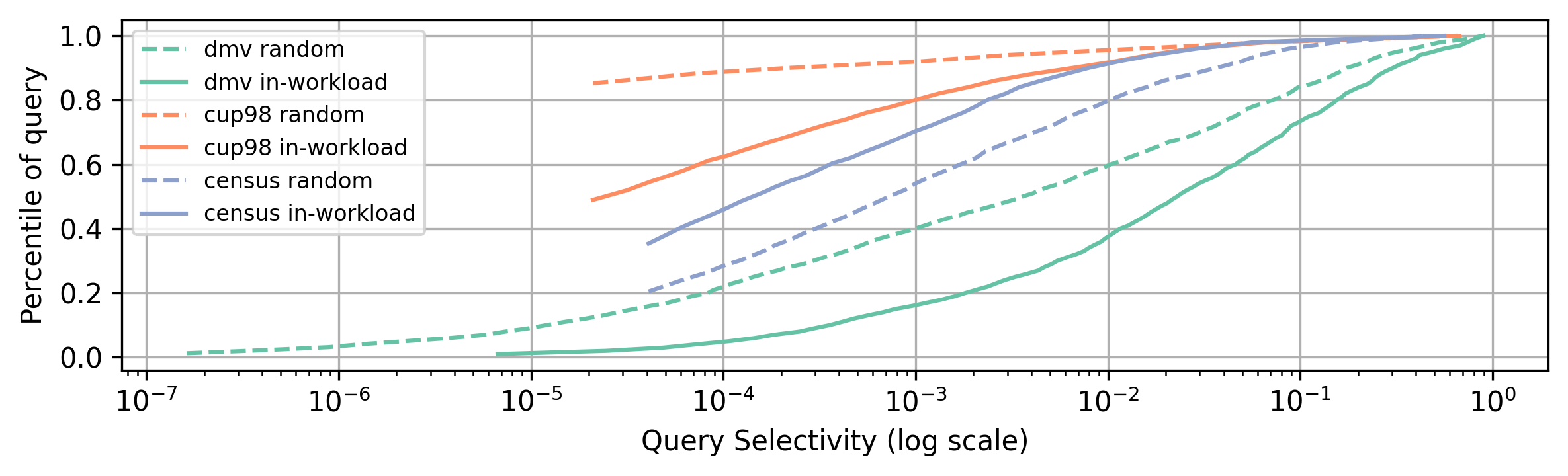}
  \vspace{-0.5cm}
  \caption{The cumulative distribution of test workloads}
  \vspace{-0.3cm}
  \label{fig:workload_dist}
\end{figure}

\textbf{Workload Cardinality Distribution.} \autoref{fig:workload_dist} plots the cardinality distribution of all these generated workloads. We can observe that the distribution between training workloads (in-workload) and random workloads is significantly different. By comparing the results between them, we can prove that, as a hybrid method rather than a query-driven method, Duet does not suffer from the workload drift problem at all.

\subsubsection{\textbf{Metric}} We use Q-Error~\cite{QError} to evaluate the cardinality estimation accuracy as most cardinality methods, which is effective in measuring CE problem~\cite{AreWeReady}. It is defined as follows: 
\begin{equation}
    \label{equation:Q-Error}
    \text{Q-Error} = \frac{\max{(estimate, actual)}}{\min{(estimate, actual)}}
\end{equation}
For the time and space cost, we measure the model size, convergence speed, inference memory cost, and the inference time cost of different parts.

\subsubsection{\textbf{Model Architectures}} We follow Naru and use a MADE with hidden units: 512, 256, 512, 128, 1,024 for DMV. For Kddcup98 and Census, we follow UAE and use 2 layers of ResMADE with hidden units 128 as our autoregressive models. \textbf{Note that Duet uses the MADE setting with Naru and UAE for a given dataset, this ensures our comparison is fair.} Since the time efficiency of cardinality estimation in real-world scenarios is critical, our evaluations focus on the MADE's performance. However, it is reasonable to expect that Duet can achieve much higher speed and scalability improvement on Transformer since its cost is higher for a single forward pass. 

\subsubsection{\textbf{Baselines}} We compare Duet to 7 previous cardinality estimation methods, including various methods.

\textbf{Non-learning Methods.} Although traditional non-learning methods are much less accurate, they are widely used due to their high efficiency. We compare Duet with them to evaluate Duet's time efficiency.

\begin{enumerate}[label={\arabic*.}]
    \item \textbf{Sampling.} This estimator uniformly samples $p\%$ tuples into the memory and applies estimation on those samples. We use the implementation from \cite{Naru, NaruCode}.
    \item \textbf{Independent (Indep).} This estimator assumes that each column is independent and uses the multiplication of each column's predicates' selectivity to estimate cardinality. We use the implementation in \cite{Naru, NaruCode}.
    \item \textbf{MHist~\cite{MHist1, MHist2}.} MHist is an $n$-dimensional histogram estimator which is widely used in most databases. We use the implementation in \cite{AreWeReady, AreCELearnedYet}.
\end{enumerate}
\textbf{Query-driven Models.} Query-driven models are the first kind of learned cardinality estimators. They can achieve high accuracy when the evaluation queries have the same distribution as the training data. We choose the most representative and widely used methods as competitors.
\begin{enumerate}[label={\arabic*.}]
    \setcounter{enumi}{2}
    \item \textbf{MSCN~\cite{MSCN}.} This query-driven estimator uses a multi-set convolutional neural network to deal with multiple tables, joins, and then performs an end-to-end estimation. We use MSCN (bitmaps) as the baseline, which is the most accurate MSCN type~\cite{MSCN} and the implementation is in~\cite{AreWeReady, AreCELearnedYet}.
    % and predicates. The convolutional results are average pooled, concatenated, and processed with an MLP network for the final estimation result.
\end{enumerate}
\textbf{Data-driven Models.} Data-driven models include the most advanced estimation models. We compare both Duet's procession and the cost with them.
\begin{enumerate}[label={\arabic*.}]
    \setcounter{enumi}{4}
    \item \textbf{DeepDB~\cite{DeepDB, AreCELearnedYet}.} This model learns sum-product networks (SPN) to represent the joint data distribution. This is a novel method that uses SPN, a non-neural deep model, to solve the cardinality estimation problem.
    \item \textbf{Naru~\cite{Naru}.} Naru is the first method that learns the dataset's distribution with a deep autoregressive neural network. Duet is implemented based on Naru's code. We also extend its code to support two-sided queries. Therefore, we can compare Duet with it on both single-sided and two-sided queries. We used the recommended hyper-parameter setting described in its documentation~\cite{NaruCode}. We use the same network architecture as Duet.
\end{enumerate}
\textbf{Hybrid Models}
\begin{enumerate}[label={\arabic*.}]
    \setcounter{enumi}{6}
    \item \textbf{UAE~\cite{UAECode}.} UAE is the first method that expands Naru into hybrid training. It revamps the progressive sampling process with the Gumbel-Softmax trick to make Naru's estimation process differentiable so they can apply backpropagation to it and improve the model's accuracy. We modified its code slightly to adapt our test workload. We use the same network architecture as Duet.
\end{enumerate}

\subsubsection{\textbf{Hardware}} We run our experiments on a server with 4 RTX A6000 (48GB memory) GPUs, Intel(R) Xeon(R) Silver 4210R CPU, and 504GB RAM if there is no special instruction in the corresponding subsection.

\subsubsection{\textbf{Ablation Study}} To evaluate the effectiveness of the hybrid training of Duet, we also run Duet$_{\mathcal{D}}$ which is training with data-driven only. 

\subsection{Hyper-Parameter Studies}
\label{Subsection: Hyper-Parameter Studies}
We first report the experimental results of the impact of the most important hyper-parameter during Duet's training, since the rest of the experiments are performed based on this hyper-parameter. We run experiments on the Kddcup98 dataset by training Duet with both data and training workloads, then evaluate its performance on the testing workload of \textit{Rand-Q}.

\begin{figure}[htbp]
  \centering
  \includegraphics[width=0.9\linewidth]{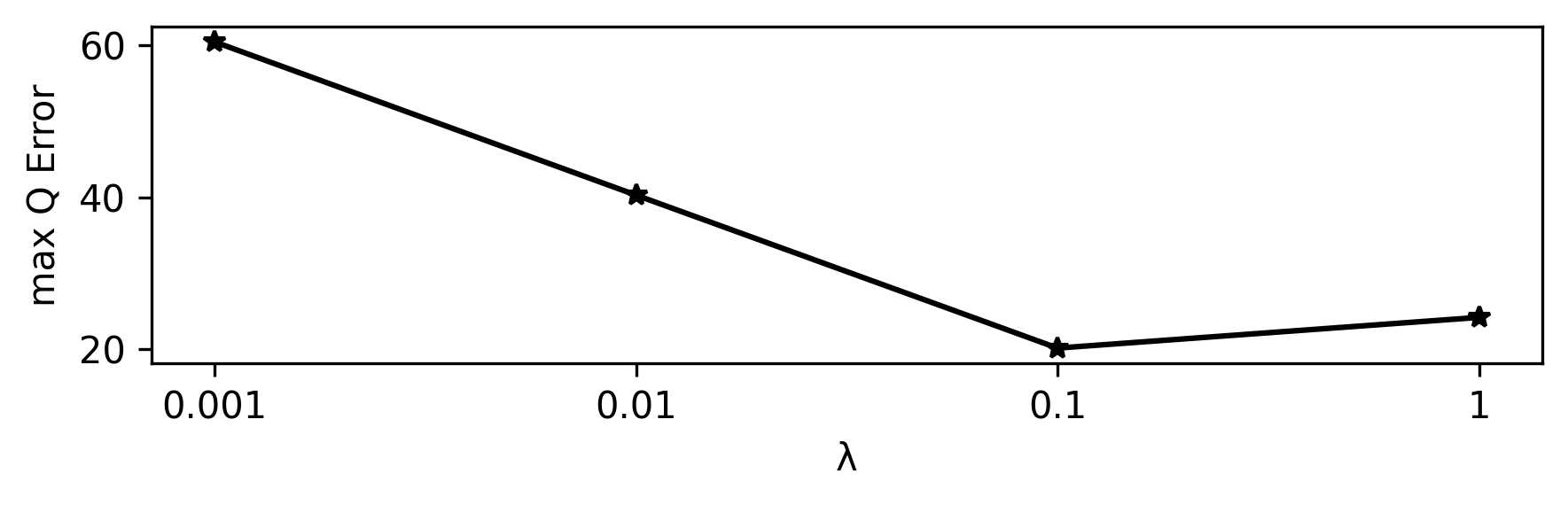}
  \vspace{-0.6cm}
  \caption{Hyper-parameter study on $\lambda$.}
  \vspace{-0.5cm}
  \label{fig:hyper-parameter_study}
\end{figure}

\textbf{Impact of trade-off coefficient $\lambda$.} The trade-off coefficient $\lambda$ affects Duet's accuracy on random queries and in-workload queries. Since we assume that Duet should be available in most hard scenarios where the users know nothing about the incoming query workload's distribution, we aim to find the $\lambda$ that makes Duet have the best generalization performance. For those scenarios where the incoming queries have a similar distribution with the training workloads, the users can increase $\lambda$ appropriately to improve the estimation accuracy. We explore the values \{$1e^{-3}, 1e^{-2}, 1e^{-1}, 1$\} and report the result in \autoref{fig:hyper-parameter_study}. The result shows that $1e^{-1}$ is the best value for $\lambda$, and we take this value in the rest of the experiments.

\subsection{Evaluation on Multiple Predicates Support}
To decide which MPSN we should use for the rest experiments, we evaluate the three kinds of MPSN proposed in \autoref{Section: Support Multiple Predicates on A Single Column}. 

We train Duet with dataset Census and the training workload introduced in \autoref{Subsubsection: Query Workloads}. Then we evaluate Duet with the $Rand-Q$ workloads and observe the best max Q-Error, estimation cost, training cost, and the epoch when we get the best model to evaluate each method's accuracy and efficiency.

The networks of \textit{MLP} and \textit{REC} used have 2 hidden layers with 64 hidden units, activated by ReLU. The \textit{RNN} includes a network with a 2-layer LSTM followed by a Full-Connected layer (FC layer), both of which have 64 hidden units for each layer. The different predicates' outputs of the LSTM are processed by the FC layer and summed together. Each column has an independent MPSN. The results are shown in \autoref{table:Evaluation Result for Multiple Predicates Support}. The \textit{MLP} method outperforms other methods on efficiency with the lowest cost of training and estimation. Although \textit{RNN} converges faster and has slightly better accuracy, we still choose \textit{MLP} as the default method to support multiple predicates for efficiency considerations.

\begin{table}[htp]
\vspace{-0.5cm}
\caption{Evaluation Results for Multiple Predicates Support}
\begin{center}
\begin{tabular}{lllll}
\toprule
name & max Q-Error     & est cost(ms)    & train cost(s)     & best epoch \\ \midrule
MLP  & 9.0            & \textbf{2.041}  & \textbf{31.894}   & 40          \\
REC  & 10.0           & 2.197           & 38.502            & 34          \\
RNN  & \textbf{8.0}   & 5.129           & 46.749            & \textbf{31}     \\    
\bottomrule
\end{tabular}
\label{table:Evaluation Result for Multiple Predicates Support}
\end{center}
\vspace{-0.8cm}
\end{table}

\subsection{\textbf{Scalability and Inference Cost Evaluation}}
\label{Subsection: Scalability and Inference Cost Evaluation}
We evaluate the scalability and inference cost first since it is the most important contribution.

As we have analyzed above, Duet is a sample-free cardinality estimation method that reduces the computation complexity from $O(n)$ to $O(1)$. Compared to the $O(n)$ complexity methods such as Naru and UAE, Duet only needs to infer the neural network one time per estimation. We compare Duet's scalability with them on Kddcup98. We use a model which is trained on all 100 columns and generates workloads that involve 2-100 columns. We apply evaluation on these workloads to compare these methods' scalability. 

\begin{figure}[htbp]
  \centering
  \vspace{-0.8cm}
  \includegraphics[width=\linewidth]{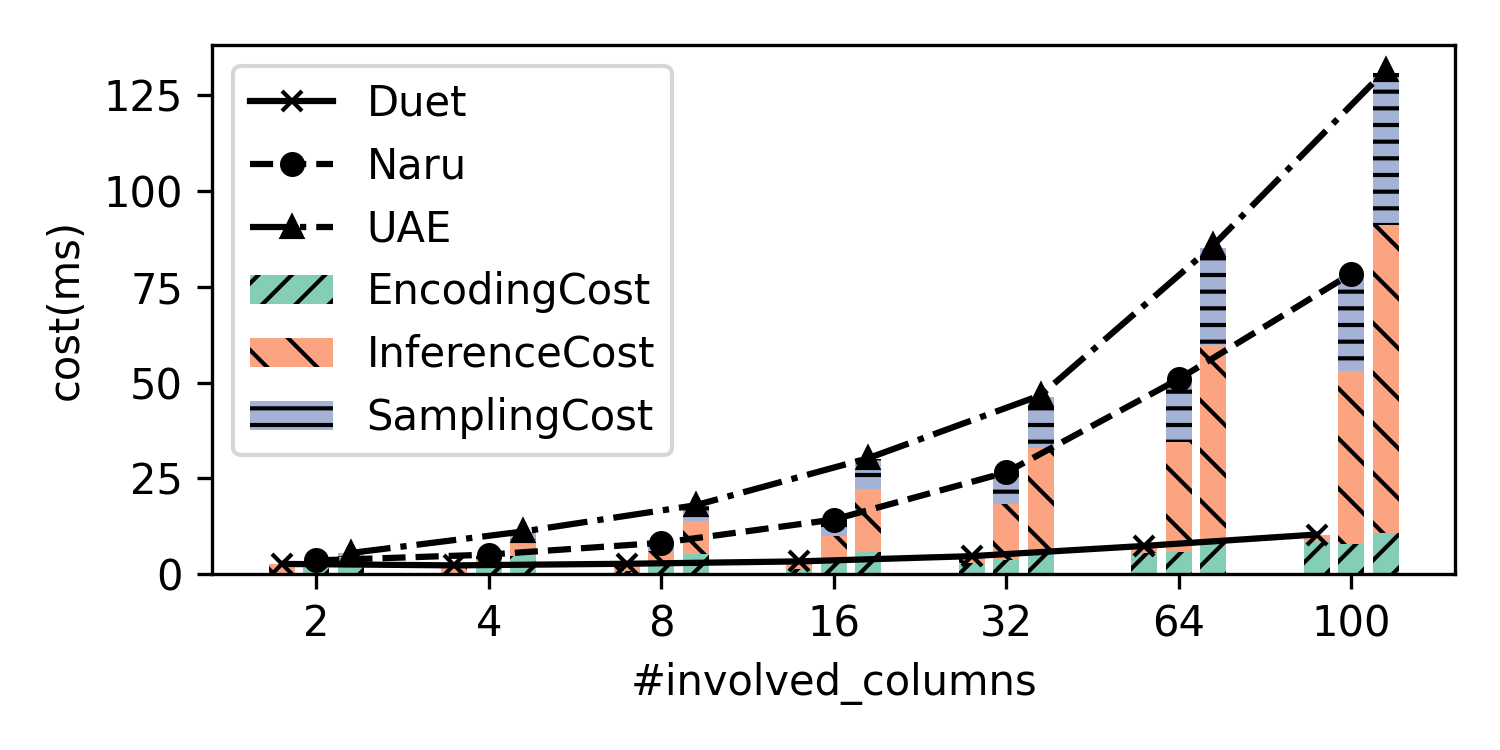}
  \vspace{-1.0cm}
  \caption{The scalability comparison on different column sizes.}
  \vspace{-0.3cm}
  \label{fig:scalability}
\end{figure}

As \autoref{fig:scalability} shows, the time cost of Naru and UAE increases with the column number linearly (the X axis is in log scale), which proves the correctness of our analysis about their $O(n)$ complexity and limited scalability. We also notice that the major increase comes from inference and sampling overhead. For Duet, it has the lowest time cost in all workloads and its increase is much slower than Naru and UAE. Also, the major increase comes from the encoding overhead. Since the encoding process is independent for each column, it can be further accelerated with a high speedup ratio by multi-threading. Besides, for the query optimization of a given query, the cardinality may be called for multiple times, but the encoding process only needs to be executed one time, and the results can also be cached. Overall, Duet is proven to have much better scalability which makes it practical and can actually help to improve the cardinality estimation accuracy for real-world scenarios.

\begin{figure*}[htbp]
  \centering
  \includegraphics[width=0.9\linewidth]{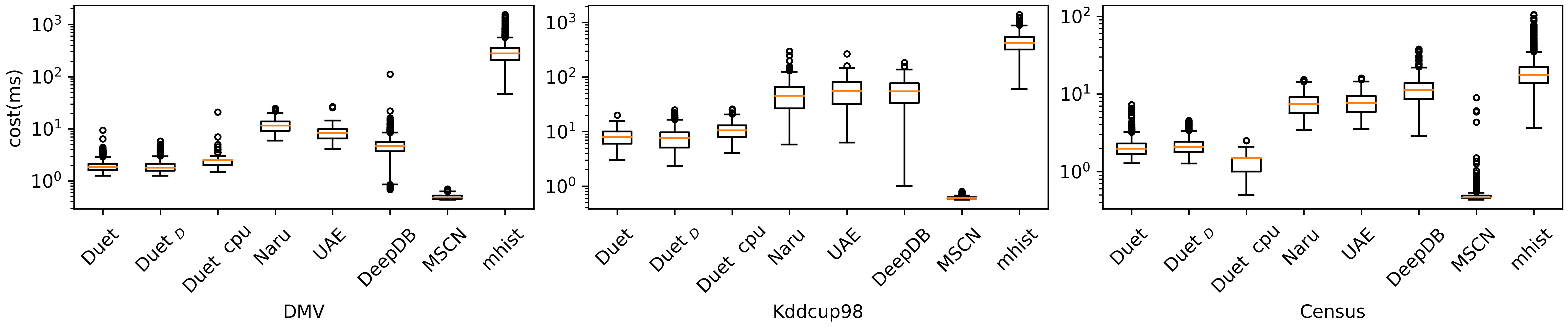}
  \vspace{-0.2cm}
  \caption{Estimation cost comparison of learned methods}
  \vspace{-0.2cm}
  \label{fig:EstimationCostComparison}
\end{figure*}

We also report the detailed inference cost of Duet and Duet$_{\mathcal{D}}$ and compare them with other learned methods since all traditional methods have too high errors. As \autoref{fig:EstimationCostComparison} shown, Duet's estimation cost is significantly lower than others, which proves that Duet's efficiency is improved by learning from the virtual table and estimating without sampling and multiple forward passes. By comparing Duet's cost on CPU with others in \autoref{fig:EstimationCostComparison}, we can draw the conclusion that \textbf{Duet even has a lower inference cost on CPU than that of both previous data-driven and hybrid methods on GPU}, so it is much cheaper to deploy Duet on real-world DBMS cluster than previous work. We also notice that MSCN has the lowest estimation cost among all learned methods, this is because MSCN uses a simple end-to-end network. However, we prove that such query-driven method has much higher error than data-driven and hybrid methods in \autoref{Subsection: Accuracy Comparison}.

\textbf{Overall, all experiments significantly prove that we have achieved our design goal on efficiency and scalability and solved the Problem (1) introduced in \autoref{Section: introduction}.}

\subsection{\textbf{Accuracy Comparison}}
\label{Subsection: Accuracy Comparison}
\begin{table*}[htbp]
\caption{Accuracy Result Of All Methods on Three Datasets}
\begin{center}
\scalebox{0.85}{
\begin{tabular}{|l|l|l|l|lllll|lllll|}
\hline
\multirow{2}{*}{Dataset}&\multirow{2}{*}{Estimator} & \multirow{2}{*}{Size(MB)} & \multirow{2}{*}{Cost(ms)} & \multicolumn{5}{c|}{In-Workload Queries}                                                          & \multicolumn{5}{c|}{Random Queries}        \\ \hhline{|~|~|~|~|-----|-----|}
                        &                           &                           &                           & mean            & median          & 75th           & 99th           &  max           & mean           & median         & 75th          & 99th           & max  \\ \hhline{:=:=:=:=:=====:=====:}
\multirow{9}{*}{DMV}    &Sampling                   & 9.8                       & 0.29                      & 17.74           & 1.027           & 1.093          & 685.5          & 2224           & 46.35          & 1.081          & 1.639         & 1082           & 1776\\ 
                        &Indep                      & -                         & 290.0                     & 79.06           & 1.257           & 1.540          & 1392.4         & 17120          & 71.14          & 1.276          & 1.624         & 685.8          & 45437\\
                        &MHist                      & 17.0                      & 221.1                     & 28.06           & 2.700           & 5.733          & 529.0          & 4409.8         & 8.905          & 2.181          & 4.142         & 102.8          & 2727.0 \\ \hhline{|~|=:=:=:=====:=====|}  
                        &MSCN                       & 6.12                      & 0.49                      & 2.783           & 1.056           & 1.233          & 28.84          & 386.0          & 11.92          & 1.406          & 3.312         & 231.4          & 746.0    \\ \hhline{|~|=:=:=:=====:=====|}
                        &DeepDB                     & 0.165                     & 4.94                      & 5.573           & 1.078           & 1.201          & 18.23          & 4191.0         & 2.619          & 1.100          & 1.254         & 20.06          & 414.0    \\
                        &Naru                       & 15.5                      & 11.4                      & \textbf{1.027}  & \textbf{1.009}  & \textbf{1.021} & \textbf{1.260} & \textbf{3.0}   & \textbf{1.064} & \textbf{1.015} & \textbf{1.037}& \textbf{2.0}   & \textbf{11.0}   \\ \hhline{|~|=:=:=:=====:=====|}
                        &UAE                        & 2.5                       & 7.94                      & 1.080           & 1.035           & 1.073          & 1.869          & 7.871          & 1.119          & 1.032          & 1.077         & 2.358          & 14.83    \\ \hhline{|~|=:=:=:=====:=====|}
                        &Duet$_{\mathcal{D}}$       & 16.3                      & 1.77                      & 1.081           & 1.022           & 1.058          & 2.390          & 4.489          & 1.107          & 1.026          & 1.072         & 2.445          & \textbf{11.0}    \\
                        &Duet                       & 16.3                      & 1.84                      & 1.050           & 1.010           & 1.025          & 1.692          & 3.671          & 1.085          & 1.016          & 1.050         & 2.401          & \textbf{11.0}    \\ \hhline{:=:=:=:=:=====:=====:}

\multirow{9}{*}{Kddcup98}&Sampling                  & 3.2                       & 0.26                      & 8.384           & 1.219           & 4.0            & 103.1          & 271.0          & 3.464          & \textbf{1.0}   & \textbf{1.0}  & 101.0          & 345.0     \\
                        &Indep                      & -                         & 9.20                      & 2.599           & 1.133           & 2.0            & 13.01          & 584.0          & 1.233          & \textbf{1.0}   & \textbf{1.0}  & 5.0            & 69.0   \\
                        &MHist                      & 45.5                      & 455.3                     & 13.46           & 2.0             & 8.0            & 167.7          & 378.0          & 4.032          & \textbf{1.0}   & \textbf{1.0}  & 65.01          & 606.0  \\ \hhline{|~|=:=:=:=====:=====|}  
                        &MSCN                       & 7.07                      & 0.62                      & 3.674           & 2.0             & 4.0            & 29.00          & 111.0          & 3.746          & 3.0            & 4.0           & 21.51          & 115.0     \\ \hhline{|~|=:=:=:=====:=====|}  
                        &DeepDB                     & 2.7                       & 39.1                      & 1.574           & 1.064           & 1.500          & 6.505          & 160.8          & 1.117          & \textbf{1.0}   & \textbf{1.0}  & 3.125          & 19.0   \\
                        &Naru                       & 3.4                       & 32.4                      & 1.500           & 1.070           & 1.500          & 5.0            & 97.33          & 1.122          & \textbf{1.0}   & \textbf{1.0}  & 2.667          & 69.0     \\ \hhline{|~|=:=:=:=====:=====|}  
                        &UAE                        & 3.4                       & 8.41                      & -               & -               & -              & -              & -              & -              & -              & -             & -              & -      \\ \hhline{|~|=:=:=:=====:=====|}  
                        &Duet$_{\mathcal{D}}$       & 7.3                       & 6.58                      & \textbf{1.329}  & \textbf{1.056}  & \textbf{1.400} & \textbf{4.0}   & \textbf{7.584} & \textbf{1.080} & \textbf{1.0}   & \textbf{1.0}  & 2.668          & \textbf{6.0}   \\
                        &Duet                       & 7.3                       & 6.31                      & \textbf{1.312}  &	\textbf{1.0462} & \textbf{1.365} & \textbf{4.0}   & \textbf{7.0}   & \textbf{1.083} & \textbf{1.0}   & \textbf{1.0}  & \textbf{2.502} & \textbf{6.0}   \\ \hhline{:=:=:=:=:=====:=====:}

\multirow{9}{*}{Census} &Sampling                   & 0.3                       & 0.05                      & 11.30           & 2.0             & 7.0            & 150.1          & 388.0          & 12.99          & 1.975          & 8.0            & 165.1         & 417.0      \\
                        &Indep                      & -                         & 1.48                      & 5.688           & 2.0             & 5.072          & 59.0           & 234.0          & 5.139          & 2.0            & 4.500          & 49.0          & 275.0         \\
                        &MHist                      & 3.5                       & 19.2                      & 5.065           & 2.0             & 4.500          & 46.44          & 231.0          & 5.576          & 2.186          & 4.574          & 49.0          & 584.0              \\ \hhline{|~|=:=:=:=====:=====|}
                        &MSCN                       & 6.15                      & 0.54                      & 2.667           & 1.595           & 3.0            & 14.0           & 38.33          & 2.774          & 1.667          & 3.0            & 17.01         & 34.0   \\ \hhline{|~|=:=:=:=====:=====|}
                        &DeepDB                     & 0.4                       & 11.1                      & 1.872           & 1.333           & 2.0            & 8.0            & 46.0           & 2.018          & 1.360          & 2.0            & 11.0          & 47.0     \\
                        &Naru                       & 0.5                       & 8.45                      & \textbf{1.319}   & \textbf{1.114}  & \textbf{1.420} & \textbf{3.500} & \textbf{9.0}   & \textbf{1.275} & \textbf{1.117} & \textbf{1.325} & \textbf{3.0}  & \textbf{5.0} \\ \hhline{|~|=:=:=:=====:=====|}
                        &UAE                        & 0.5                       & 8.91                      & 1.357           & 1.141           & 1.500          & 3.670          & \textbf{9.0}   & 1.326          & 1.143          & 1.398          & 3.376         & \textbf{5.0} \\ \hhline{|~|=:=:=:=====:=====|}
                        &Duet$_{\mathcal{D}}$       & 0.8                       & 2.03                      & 1.477           & 1.195           & 1.641          & 4.501          & 9.667          & 1.375          & 1.155          & 1.500          & 3.751         & 7.0      \\
                        &Duet                       & 0.8                       & 2.06                      & 1.891           & 1.465           & 2.027          & 8.500          & 12.33          & 1.464          & 1.220          & 1.600          & 4.001         & 9.0    \\ \hline

\end{tabular}
}
\end{center}
\vspace{-0.8cm}
\label{table:Multiple_Accuracy}
\end{table*}

We use the exact same settings of Duet and baselines with \autoref{Subsection: Scalability and Inference Cost Evaluation} to evaluate the estimation accuracy and compare it with the baselines. We modify the implementation of the previous works~\cite{Naru, AreWeReady, UAE} to support our workloads and run them with the best setting introduced within their documentation. All data-driven and hybrid methods learn from the same datasets, and all query-driven and hybrid methods use the same training workload introduced in \autoref{Subsubsection: Query Workloads}. We evaluate Duet and Duet$_{\mathcal{D}}$ on both \textit{Rand-Q} and \textit{In-Q} workloads to compare their accuracy. The \autoref{table:Multiple_Accuracy} shows the results of all methods' accuracy on different workloads and settings, we can obtain a few conclusions from them.
\begin{itemize}
    \item \textbf{Duet outperforms all methods on the high-dimensional table with both test workloads.} We draw this conclusion from the results of Kddcup98. Both Duet and Duet$_{\mathcal{D}}$ outperform all methods which shows that we can obtain huge accuracy improvement by removing sampling.
    
    \item \textbf{Hybrid training can improve Duet's accuracy on both In-workload and random queries for high-cardinality and high-dimensional tables.} The ablation study of Duet's hybrid training and data-driven training on DMV and Kddcup98 helps us to draw this conclusion. Note that due to its loss design, UAE has a gradient explosion problem caused by excessive $\mathcal{L}_{query}$ during training on Kddcup98, which cannot be fixed by loss clamping. For the random queries, the Duet with hybrid training also shows an accuracy improvement. But on the Census, the hybrid training seems to slightly lower its accuracy. We believe that this is because the Census dataset is so simple that the hybrid training only makes the model harder to converge since the Q-Error is quite unstable when the training just starting.
    
    \item \textbf{Duet does not suffer from the workload drift problem.} Considering Duet's results of Random Queries don't drop significantly compared to In-Workload Queries, and the huge difference between the two workloads, we can also draw this conclusion since it learns from the data mostly, and the query information is just used as a supplement.

    \item \textbf{Overall, Duet's accuracy is about the same as Naru and other hybrid methods represented by UAE in other experiments.} We observe from \autoref{table:Multiple_Accuracy} that Naru and UAE's accuracy is slightly higher than Duet's. This is because the Census dataset is the smallest, so the shortcomings of their progressive sampling are not significant. And for the DMV dataset, the high number of distinct values leads to a huge sampling space. Since Duet learns from the samples of the virtual table, it is hard for it to be trained on every possible tuple. The hybrid training can further improve Duet's accuracy in this situation, but since we use training workloads with bounded predicates, the improvement is limited. By removing the limitation of the bounded column, we observe that Duet achieves a significantly lower max Q-Error than Naru and UAE on In-workload queries(2.312 vs 14.0).
    
    \item \textbf{Duet alleviates the long-tail distribution problem.} According to results on Kddcup98, removing sampling reduces the max Q-Error over 10 times and solves the long-tail problem for high-dimensional tables.
    
    %Comparing the performance between Duet and Duet$_{\mathcal{D}}$ in DMV and Kddcup98, Duet that is trained with both data and workloads achieves better accuracy compared to Duet$_{\mathcal{D}}$ in both high cardinality and high-dimensional situations, especially on In-workload queries. Considering that hybrid training takes a higher cost, we suggest disabling it for small tables.    
\end{itemize}

\textbf{In general, the accuracy evaluation proves that we have solved the Problem (2) introduced in \autoref{Section: introduction}.}

\subsection{\textbf{Training Cost Comparison}}
We use the same setting with \autoref{Subsection: Accuracy Comparison} to train Duet and Duet$_{\mathcal{D}}$, and compare its convergence speed, throughput, and GPU memory cost with Naru and UAE.

\begin{figure}[htbp]
  \centering
  \includegraphics[width=\linewidth]{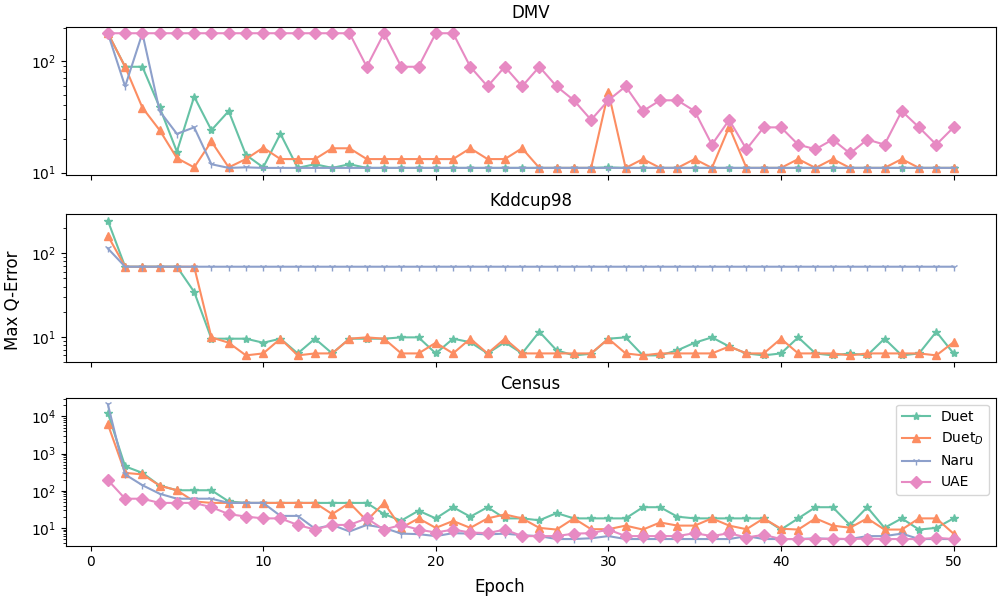}
  \vspace{-0.8cm}
  \caption{Convergence speed of different learned methods on \textit{Rand-Q}.}
  \vspace{-0.3cm}
  \label{fig:ConvergenceSpeedCompare}
\end{figure}

\autoref{fig:ConvergenceSpeedCompare} shows the convergence of the max Q-Error as training progresses when evaluated on random queries. The results indicate that Duet converges significantly faster and better than Naru and UAE on the high-dimensional dataset (Kddcup98). Even on the biggest dataset DMV, Duet still takes fewer epochs to achieve the same max Q-Error than Naru. This is because Duet can not only learn each data point's probability but also learn the actual semantics of predicates. Therefore, in the iterative training process, since the sampled virtual tuples used in training Duet are generated from all tuples, Duet can still converge at a fast speed. This also proves the effectiveness of our sampling approach~\autoref{Algorithm: Virtual Tuple Sampling Method}. Also, we observe that the UAE converges much slower than others on the DMV dataset, The reason is that the DMV dataset has higher cardinality, and since UAE just scales $\mathcal{L}_{query}$ with a single factor, the initial high value of the $\mathcal{L}_{query}$ (shown in \autoref{fig:loss_convergence}) affects its convergence. This proves the effectiveness of our hybrid training loss.

\begin{figure}[htbp]
  \centering
  \includegraphics[width=\linewidth]{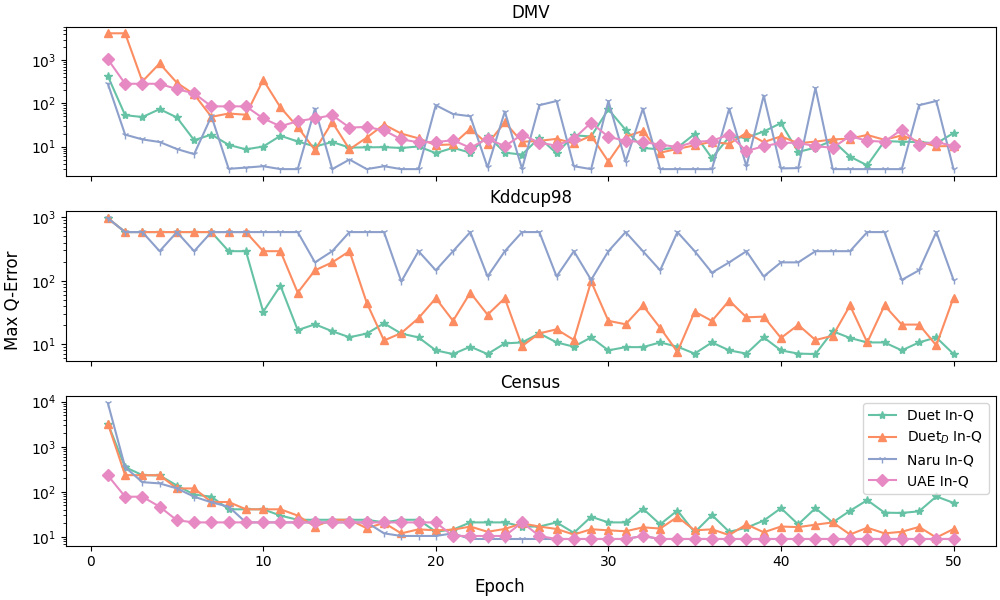}
  \vspace{-0.8cm}
  \caption{Convergence speed of different learned methods on \textit{In-Q}.}
  % \vspace{-0.3cm}
  \label{fig:ConvergenceSpeedCompareIn}
\end{figure}

By comparing with Duet and Duet$_{\mathcal{D}}$ in \autoref{fig:ConvergenceSpeedCompareIn}, we draw a conclusion that the hybrid training slightly increases the convergence speed when evaluated on in-workload queries.


Since Duet requires sampling from the virtual table during training, it is reasonable to question its training efficiency. We test its training throughput and compare it with the throughput of Naru and UAE. Since GPU devices like A6000 are too expensive to be ensembled with DBMS clusters, we use a PC with RTX3080(10GB) and AMD Ryzen9 5900X to evaluate the training throughput. The batch size is set to 2,048 for DMV and 100 for Kddcup98 and Census followed by UAE's setting.

The result shown in \autoref{table:Throughput Comparison} indicates that Duet's throughput is $42.5\%$ lower than Naru and $556.9\%$ higher than UAE on DMV. For memory cost, Duet costs 3.5GB GPU memory for training DMV, Duet$_{\mathcal{D}}$ costs 2.2 GB, Naru costs 2.1 GB, and UAE costs 2.8 GB. As a hybrid method. We emphasize that the $\#sample$ of UAE and Naru during evaluation is 2,000, if we set it to 2,000 during training, the UAE will cost over 10GB GPU memory and trigger an OOM exception on RTX3080. Even with $\#sample$ set to 200, it still triggers the OOM exception on the Kddcup98 dataset. The accuracy evaluation in \autoref{Subsection: Accuracy Comparison} indicates that this is a part of the reasons why UAE achieves lower accuracy on in-workload queries than Duet. \textbf{Thus, Duet has much better time and memory efficiency for training and updating than UAE in general, and Problem (3) is also solved.}
\vspace{-0.5cm}
\begin{table}[htbp]
\caption{Throughput (tuples/s) of data-driven and hybrid methods.}
\begin{center}
\begin{tabular}{llll}
\toprule
Estimator            &  DMV    & Kddcup98 & Census\\
\midrule
Naru                 &188006.4 &  2963    & 14538  \\
UAE                  &16445.4  &  OOM     & 532    \\
Duet$_{\mathcal{D}}$ &160358.4 &  1404    & 8332   \\
Duet                 &108032   &  1019    & 5516    \\
\bottomrule
\end{tabular}
\label{table:Throughput Comparison}
\end{center}
\vspace{-0.8cm}
\end{table}

\section{Conclusion and Future Works}
We propose a novel methodology for modeling data-driven cardinality estimation avoiding sampling for range queries and any non-differentiable process during inference. Based on this methodology, we propose Duet, a stable, efficient, and scalable hybrid cardinality estimation method that reduces the inference complexity from $O(n)$ to $O(1)$ and has less memory cost for training than Naru and UAE. The experimental results also show that Duet solves Problems (1-3) introduced in \autoref{Section: introduction}. In terms of inference cost, Duet can even achieve faster estimation on CPU than that of most learned methods on GPU. And since Duet is a deterministic method and mainly learns from data, the instability problem and workload drift problem (Problems (4,5)) are avoided. In summary, we have proven that Duet is more practical for real-world scenarios compared to previously learned methods such as MSCN, DeepDB, Naru, and UAE, and can help drive the practical implementation of learning-based cardinality estimators. We aim to improve the performance of Duet on tables with extremely large distinct values in our future work.

\clearpage

% \section*{Acknowledgment}

% The preferred spelling of the word ``acknowledgment'' in America is without 
% an ``e'' after the ``g''. Avoid the stilted expression ``one of us (R. B. 
% G.) thanks $\ldots$''. Instead, try ``R. B. G. thanks$\ldots$''. Put sponsor 
% acknowledgments in the unnumbered footnote on the first page.

\vspace{12pt}
% IEEE conference templates contain guidance text for composing and formatting conference papers. Please ensure that all template text is removed from your conference paper prior to submission to the conference. Failure to remove the template text from your paper may result in your paper not being published.

\end{document}